\documentclass[twoside,slac_one]{revtex4}
\usepackage{graphicx}
\usepackage{fancyhdr}
\usepackage{amsmath} 
\usepackage{bm}
\usepackage{amsxtra}
\usepackage{amssymb}
\usepackage{amsthm}
\usepackage{latexsym}
\usepackage{lscape}

\begin{document}

\global\long\def\met{\not{\!{\rm E}}_{T}}

\title{LHC Sensitivity to $Wb\bar{b}$ Production via Double Parton Scattering}

%

\author{Seth Quackenbush}
\author{Edmond L.~Berger}
\affiliation{High Energy Physics Division, Argonne National Laboratory, Argonne, IL 60439}
\author{C.~B.~Jackson}
\affiliation{Physics Department, University of Texas at Arlington, Arlington, TX 76019}
\author{Gabe Shaughnessy}
\affiliation{High Energy Physics Division, Argonne National Laboratory, Argonne, IL 60439 \\ Department of Physics and Astronomy, Northwestern University, Evanston, IL 60208}

\begin{abstract}
We investigate the potential to observe double parton scattering at the Large Hadron Collider  in 
$p p \rightarrow Wb\bar{b} X \to \ell \nu b \bar{b} X$ at 7 TeV.  Our analysis tests the efficacy of several kinematic variables in isolating the double parton process of interest from the single parton process and relevant backgrounds for the first 10 fb$^{-1}$ of integrated luminosity.  These variables are constructed to expose the independent nature of the two subprocesses in double parton scattering, $pp \to \ell \nu X$ and $pp \to b \bar{b} X$.  
We use next-to-leading order perturbative predictions for the double parton and single parton scattering components of $W b \bar{b}$ and for the pertinent  backgrounds.   The next-to-leading order contributions are important for a proper description of some of the observables we compute.  We find that the double parton process can be identified and measured with significance $S/\sqrt B \sim 10$, 
provided the double parton scattering effective cross section $\sigma_{\rm eff} \sim 12$~mb.  
\end{abstract}

\maketitle

\thispagestyle{fancy}


\section{Introduction}
\label{sec:intro}

The successful operation of the Large Hadron Collider (LHC) and its detectors opens a new era in particle physics.  The higher energies and larger luminosities at the LHC make it possible to explore new physics scenarios and to investigate unexplored aspects of established theories such as quantum chromodynamics (QCD).

\begin{figure*}[ht]
\includegraphics[width=60mm]{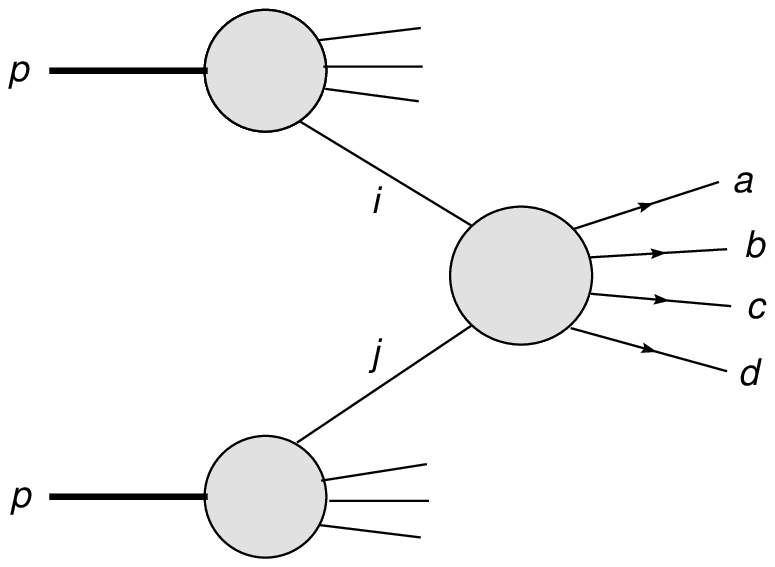}
\hspace{1.5cm}
\includegraphics[width=60mm]{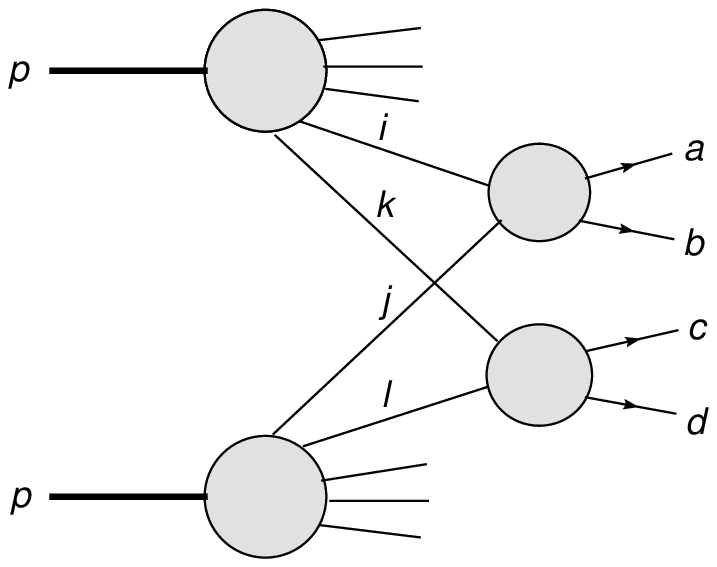}
\caption{Schematic depiction of single parton scattering (left) and double parton scattering (right).  Labels are as in the text. }
\label{fg:sps-dps-cartoons}
\end{figure*}

The standard picture of hadron-hadron collisions is shown on the left side of Fig.~\ref{fg:sps-dps-cartoons}.  One parton from each proton partakes in the hard scattering to produce the final state.  The probability density for finding parton $i$ in a proton with momentum fraction $x_i$ and at the factorization scale $\mu$ is parameterized by the parton distribution function (PDF) $f^i_p(x_i, \mu)$.  In this {\it single parton scattering} (SPS) scenario, the differential hadronic cross section neatly factors into: 
\begin{equation}
d\sigma^{SPS}_{pp} = \sum_{i,j} \int f^i_p(x_1,\mu) f^j_p(x_1^\prime, \mu) d\hat{\sigma}_{ij}(x_1,x_1^\prime,\mu) dx_1 dx_1^\prime \,.  
\label{eq:sps-xs}
\end{equation}
The ``short-distance" partonic cross section $d\hat{\sigma}_{ij}$ is computed in perturbation theory, whereas the PDFs are non-perturbative objects and must be extracted from experiment.

The full description of hadronic collisions involves other elements including initial- and final-state soft radiation,  underlying events, and multi-parton interactions.  Double parton scattering (DPS) describes the case in which two short-distance subprocesses occur in a given hadronic interaction, with two initial partons being active from each of the incident protons.  The general picture of DPS is shown on the right side of Fig.~\ref{fg:sps-dps-cartoons}.     Some evidence for DPS has been observed at the CERN Intersecting Storage Rings \cite{Akesson:1986iv}, the CERN SPS \cite{Alitti:1991rd},  and more recently, at the Fermilab Tevatron \cite{Abe:1997xk,D0:2009}.    

Under the assumption of weak dynamic and kinematic correlations between the two hard scattering subprocesses, the general approach in these studies is to assume the differential hadronic cross section takes a factored form in analogy to Eq.~(\ref{eq:sps-xs}):
\begin{eqnarray}
d\sigma^{DPS}_{pp} &=& \frac{m}{2 \sigma_{\rm eff}} \sum_{i,j,k,l} \int H_p^{ik}(x_1,x_2,\mu_A,\mu_B) H_p^{jl}(x_1^\prime,x_2^\prime,\mu_A,\mu_B) \nonumber\\
&& \times d\hat{\sigma}_{ij}(x_1,x_1^\prime,\mu_A) d\hat{\sigma}_{kl}(x_2,x_2^\prime,\mu_B) dx_1 dx_2  dx_1^\prime dx_2^\prime \,,
\label{eq:dps-xs}
\end{eqnarray}
where $m$ is a symmetry factor which is equal to 1 (2) if the two hard-scattering subprocesses are identical (non-identical).   The {\it joint probabilities} $H_p^{i,k}(x_1,x_2,\mu_A,\mu_B)$  can be approximated as the product of two single PDFs:
\begin{equation}
H_p^{i,k}(x_1,x_2,\mu_A,\mu_B) = f_p^i (x_1,\mu_A) f_p^k(x_2,\mu_B) \,.
\label{eq:joint-PDF}
\end{equation}
Given that one hard-scattering has taken place, the parameter $\sigma_{\rm eff}$ measures the size of the partonic core.  Typical values in phenomenological studies focus on the 10-12 mb region, consistent with measurements from the Tevatron collider~\cite{Abe:1997xk,D0:2009}.

In an earlier study~\cite{Berger:2009cm}, we investigated the DPS and SPS contributions at the LHC to the four-parton final state $p p \rightarrow b \bar{b} j j X$ in which  a $b \bar{b}$ system is produced along with two jets $j$.   We showed that there are characteristic regions of phase space in which the DPS events are expected to concentrate, and we developed a methodology to measure the effective size of DPS.  Precise measurements of DPS at the LHC will provide insight into parton correlations, non-perturbative dynamics in hadron-hadron collisions, the structure of the proton, and parton distribution functions.

In this talk, we examine another final state besides $b\bar{b}jj$ which may be a good candidate to observe DPS, namely the production of a $W$ boson in association with a pair of bottom quark jets.  The cross sections for both subprocesses, $W$ and $b\bar{b}$, are individually large, and the charged lepton from the $W$ decay (along with the bottom quarks in the final state) provides a relatively clean signal to tag on.   Our purpose is to establish whether double parton scattering can be 
observed as a discernible physics process in $Wb\bar{b}$ production at LHC energies.  We remark that the $Wb\bar{b}$ final state is a significant background for the production of a Higgs boson $H$ in the $HW^\pm$ mode, where the Higgs boson decays into a pair of bottom quarks \cite{DelFabbro:1999tf}, and that it can be a background in channels where new physics may arise such as in single top quark production~\cite{Cao:2007ea}.  Once DPS production of  $Wb\bar{b}$ is observed, it would be interesting to assess its potential significance as a background in such searches.

The rest of the talk is structured as follows.  In Section \ref{sec:procedure}, we outline our procedure for computing the DPS and SPS contributions to $Wb\bar{b}$ at the LHC, and we discuss and evaluate backgrounds to the same final state.  Section \ref{sec:ttbar-bkgd} is devoted to the role of the large $t \bar{t}$ background.  In Section \ref{sec:DPS-vs-SPS}, we focus on discrimination between the DPS and SPS contributions to $Wb\bar{b}$.  We study various single variable and two-dimensional kinematic distributions to bring out the DPS contribution more cleanly.  Section~\ref{sec:conclusions} contains our summary.

\section{DPS and SPS contributions to $Wb\bar{b}$ production at the LHC}
\label{sec:procedure}

We begin with the premise that there are DPS and SPS components of the same $Wb\bar{b}$ final state.  Our aim is to try to pick out the DPS component and to study its distinct properties.  Here we outline our method for computing event rates for $Wb\bar{b}$ from DPS and SPS as well as backgrounds at the LHC.  We perform all calculations at a center-of-mass energy of $\sqrt{s} = 7$ TeV.  Event rates are quoted for 10 fb$^{-1}$ of integrated luminosity.  For the DPS case, $Wb\bar{b}$ production is computed using Eq.~(\ref{eq:dps-xs}) where it is assumed that one hard scattering produces the $W$ boson via the Drell-Yan mechanism, while the other scattering produces the $b\bar{b}$ system.  Schematically, we can represent the partonic DPS process as: 
\begin{equation}
\left( ij \to W^\pm \right) \otimes \left( k l \to b\bar{b} \right) \,.
\label{eq:DPS-process}
\end{equation}

The individual SPS processes which make up the DPS process in Eq.~(\ref{eq:DPS-process}) are generated using the POWHEG BOX event generator~\cite{POWHEG,Alioli:2008gx,Frixione:2007nw} which includes next-to-leading order (NLO) QCD corrections for both, plus shower emission.  The $\otimes$ symbol denotes the combination of one event from each of the $W^\pm$ and $b\bar{b}$ final states.  All events are produced using CT10 NLO PDFs~\cite{Lai:2010vv}.

In the SPS production of $Wb\bar{b}$, one hard-scattering produces the complete final state.  The events from this process are also generated using the POWHEG BOX~\cite{Oleari:2011ey} which implements the NLO calculation of Ref. \cite{Febres Cordero:2006sj}.

Many standard model processes imitate the $Wb\bar{b} \to b\bar{b} \ell \nu$ final state.  In particular, we consider contributions from the following final states:
\begin{itemize}
\item Top quark pair production $t \bar{t}$ where either ({\it{i}}) both $t$'s decay semi-leptonically (denoted by $t_\ell t_\ell$), and one of the charged leptons is missed or ({\it{ii}}) where one $t$ decays semi-leptonically while the other decays hadronically (denoted by $t_\ell t_h$) and only two jets are observed. 

\item Single top quark production ($t\bar{b}$, $\bar{t}b$, $tj$ and $\bar{t}j$ modes) where $t \to W^+ b \, (\bar{t} \to W^- \bar{b})$
\item $Wjj$, where both light jets are mistagged as a $b$ jets
\item $Wbj$ where the light jet is mistagged as a $b$ jet
\end{itemize}
Other processes were found negligible after cuts.
We compute the $Wjj$ DPS contribution using the same method discussed above for $Wb\bar{b}$, using the additional POWHEG code~\cite{Alioli:2010xa}.  The top pair~\cite{Frixione:2007nw} and single top~\cite{Alioli:2009je} SPS processes are also generated using the POWHEG BOX.  Other SPS processes are generated using MadEvent~\cite{Alwall:2007st} or ALPGEN~\cite{Mangano:2002ea}; $Wjj$ SPS is reweighted using a K-factor obtained with MCFM~\cite{Campbell:2003hd}.  

In order to avoid soft and collinear divergences in the processes that we compute using MadGraph and ALPGEN, we apply a minimal set of {\it generator-level} cuts:
\begin{eqnarray}
p_{T,j} > 15\,{\rm{GeV}} \,\, , \,\,\, |\eta_j| < 4.8 \,\, , \,\,\,  |\eta_\ell| < 2.5 \,\,, \,\,\,
p_{T,b} > 15\,{\rm{GeV}} \,\, , \,\,\, |\eta_b| < 2.5 \,\, , \,\,\,
\Delta R_{j(b)j(b)} > 0.4 \,\, , \,\,\, \Delta R_{j(b)\ell} > 0.4 \,,
\end{eqnarray}
where $\eta$ is the pseudorapidity and $\Delta R_{l k} = \sqrt{ \left( \eta_l - \eta_{k} \right)^2 + \left( \phi_l - \phi_{k} \right)^2 }$ is the separation in the azimuthal-pseudorapidity plane between the two objects $l$ and $k$.

\subsection{Simulation}

To identify the $Wb\bar{b}$ final state and reduce backgrounds, we begin with simple identification cuts on the generated event samples.  First, we consider only muonic decays of the $W$ boson ($W \to \mu \nu$).  We limit the hadronic activity in our events to include exactly two hard jets, both of which must be identified as bottom quark jets.  Finally, all events (DPS and SPS $Wb\bar{b}$ as well as backgrounds) are required to pass the following {\em acceptance} cuts:
\begin{eqnarray}
p_{T,b} \ge 20 \,{\rm{GeV}} \,,\, |\eta_b| \le 2.5 \,;\, 
20 \, {\rm{GeV}} \le p_{T,\mu} \le 50 \, {\rm{GeV}} \,, \, |\eta_\mu| < 2.1 \,;\, 
\met \ge 20 \, \rm{GeV} ;\,
\Delta R_{b\bar{b}} \ge 0.4 \,,\, \Delta R_{b\mu} \ge 0.4 \,.
\end{eqnarray}
We also apply detector resolution smearing and mistags; details can be found in our paper~\cite{Berger:2011ep}.

Table \ref{tbl:cut-results} shows the number of events from the $Wb\bar{b}$ final state (DPS and SPS) and the backgrounds both before and after the acceptance cuts are applied.   In these results and those that follow, we sum the $W^+$ and $W^-$ events.   In evaluating the DPS processes, we assume a value $\sigma_{\rm eff} \simeq 12$ mb for the effective cross section, but the goal is to motivate an empirical determination of its value at LHC energies.  The results in Table \ref{tbl:cut-results}  make it  apparent that $Wb\bar{b}$ production from SPS and the top quark pair background are the most formidable obstacles in extracting a DPS signal.

\begin{table}[ht]
\begin{center}
\caption{Numbers of events before and after the various cuts are applied for 10 fb$^{-1}$ of data.}
\begin{tabular}{|c|c|c|c|c|}
\hline
\,\,\, \textbf{Process} \,\,\,& \,\,\, \textbf{Generator-level Cuts} \,\,\, & \,\,\, \textbf{Acceptance Cuts} \,\,\,& \,\,\, $\met \le 45$ GeV \,\,\,& \,\,\, $S_{p_T}^\prime \le 0.2$ \,\,\, \tabularnewline
\hline
$W^\pm b \bar{b}$ (DPS) & 10000 & 247 & 231 & 173 \tabularnewline
\hline
$W^\pm b \bar{b}$ (SPS) & 44000 & 1142 & 569 & 114 \tabularnewline
\hline
$t\bar{t}$ & 225000 & 1428 & 290 & 13 \tabularnewline
\hline
$W^\pm jj$ (DPS)  & 476000 & 43.5 & 37.7 & 27.3 \tabularnewline
\hline
$W^\pm jj$ (SPS)  & 20300000 & 101 & 55.7 & 19.6 \tabularnewline
\hline
Single top & 20000 & 492 & 168 &  15 \tabularnewline
\hline
$W^\pm b j$ & 153000 & 152 & 53.1 & 8.2 \tabularnewline
\hline
\end{tabular}
\end{center}
\label{tbl:cut-results}
\end{table}

\section{$t\bar{t}$ Background Rejection}
\label{sec:ttbar-bkgd}

\begin{figure*}[ht]
\includegraphics[width=65mm]{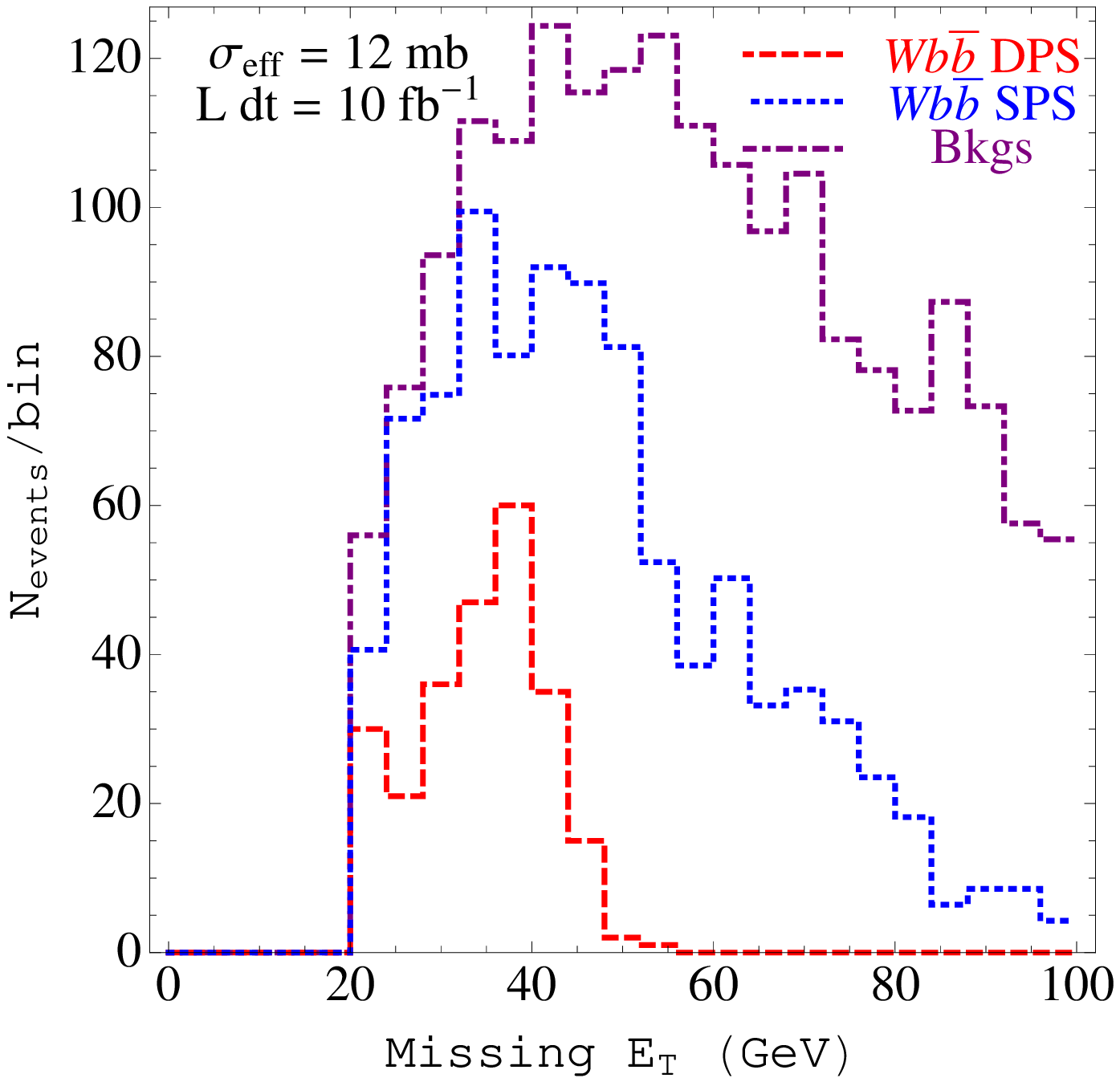}
\hspace{0.5cm}
\includegraphics[width=65mm]{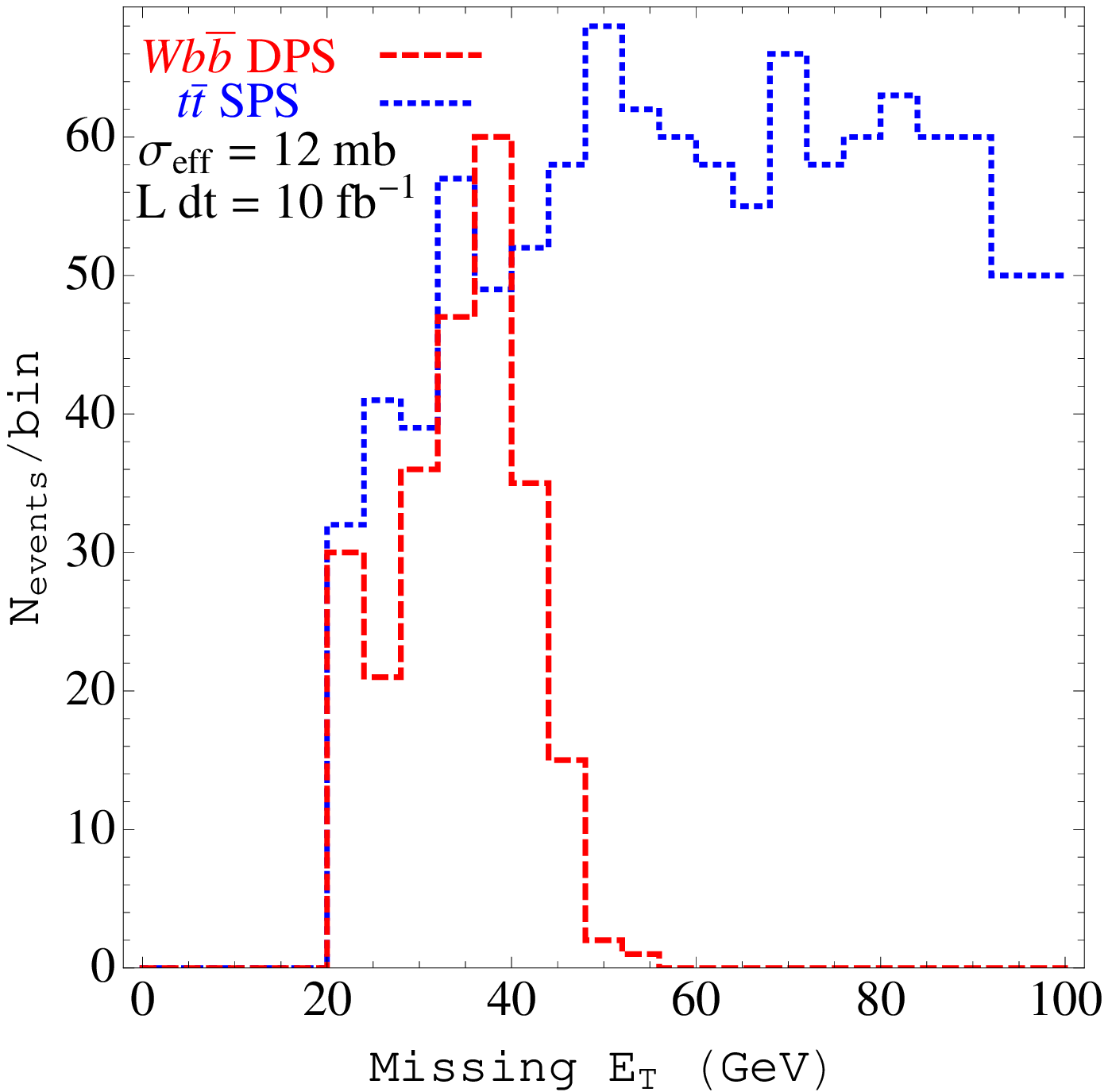}
\caption{The event rate as a function of $\met$ for DPS and SPS.  On the left, all backgrounds are included while the plot on the right compares the DPS events to those from $t\bar{t}$ alone.}
\label{fg:ET-miss}
\end{figure*}

We examine three possibilities to reduce the $t\bar{t}$ background:  a cut to restrict $\met$ from above, 
rejection of events in which a top quark mass can be reconstructed, and a cut to restrict the transverse momentum of the leading jet.    In the end, an upper cut on $\met$ in the event appears to offer the best advantage.  Indeed, one would expect that $\met$ in $Wb\bar{b}$ events would be smaller than $\met$ in $t\bar{t}$ events.  Top quark decays give rise to boosted $W^\pm$'s which, after decay, should result in larger values of missing $E_T$ compared to the $Wb\bar{b}$ process.   The $\met$ distribution is shown in Fig.~\ref{fg:ET-miss} for the DPS component of $Wb\bar{b}$, the SPS component of $Wb\bar{b}$, and {\it{all}} backgrounds (left).  On the right, we show the DPS component of $Wb\bar{b}$ and the $t\bar{t}$ background alone.  The plot on the right shows that the DPS signal is produced in the region of relatively small $\met$ and the $t\bar{t}$ background has a harder spectrum in $\met$.   One way to suppress the $t\bar{t}$ background while leaving the DPS signal unaffected is to impose a maximum $\met$ cut in the 40-60 GeV range.  In the analysis that follows, we include a maximum $\met$ cut of 45 GeV in addition to the acceptance cuts outlined above.  
The effects of the maximum $\met$ cut are shown in the fourth column of Table~\ref{tbl:cut-results}.   
This cut eliminates about 80\% of the $t\bar{t}$ background that remained after the initial acceptance cuts.

Backgrounds from events that contain a real top quark, such as the $t\bar{t}$ and single top events, might be reduced if one could reconstruct a top quark mass distribution $M_{b\ell\nu}$ 
from the final state objects (bottom quarks, charged leptons and neutrinos), and then eliminate events in 
which the reconstructed mass falls in a narrow window centered on the known top quark mass.  We attempted this by asserting the reconstruction of the $W$ mass to determine the longitudinal component of the neutrino momentum, however the $t\bar{t}$ background was not sufficiently peaked due to several issues.  These are discussed in ~\cite{Berger:2011ep}.

We also considered an upper cut on the $p_T$ of the leading object, since this would tend to be harder for jets coming from a top quark.  While the backgrounds did in fact have a larger $p_T$ tail, they are also peaked on the low end where DPS events are located, and so we abandoned this observable as a possible discriminator.

We conclude this section with the statement that of the three possibilities to reduce the $t\bar{t}$ background we considered, a cut to restrict $\met$ from above appears to offer the best advantage and is the only added cut we impose in addition to the acceptance cuts specified above.

\section{Discrimination between DPS and SPS contributions to $Wb\bar{b}$ }
\label{sec:DPS-vs-SPS}

To separate the DPS events from those of SPS origin, we find it convenient to employ quantities which take into account information from the entire final state.  One useful observable is $S^\prime_{p_T}$, defined as \cite{D0:2009}:
\begin{equation}
S^\prime_{p_T} = \frac{1}{\sqrt{2}} \sqrt{ \left( \frac{|p_T(b_1,b_2)|}{|p_T(b_1)| + |p_T(b_2)|}\right)^2 + \left( \frac{|p_T(\ell,\nu)|}{|p_T(\ell)| + |p_T(\nu)|}\right)^2 } \,.
\label{eq:Sptprime}
\end{equation}
Here $p_T(b_1,b_2)$ is the vector sum of the transverse momenta of the two $b$ jets, and 
$p_T(\ell,\nu)$ is the vector sum of the transverse momentum of the lepton and missing energy in the final state.  
\begin{figure*}[ht]
\includegraphics[width=65mm]{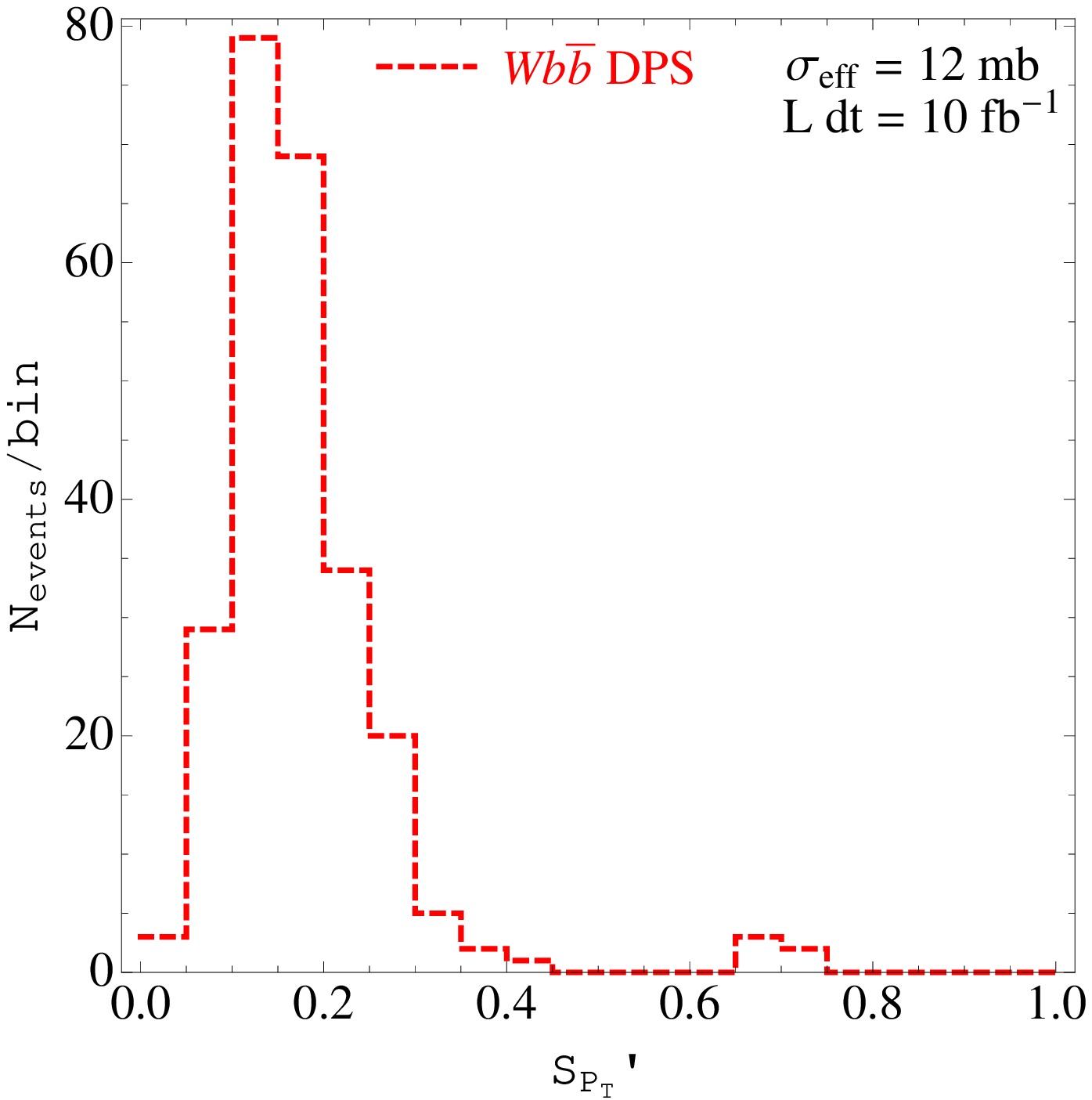}
\hspace{0.5cm}
\includegraphics[width=65mm]{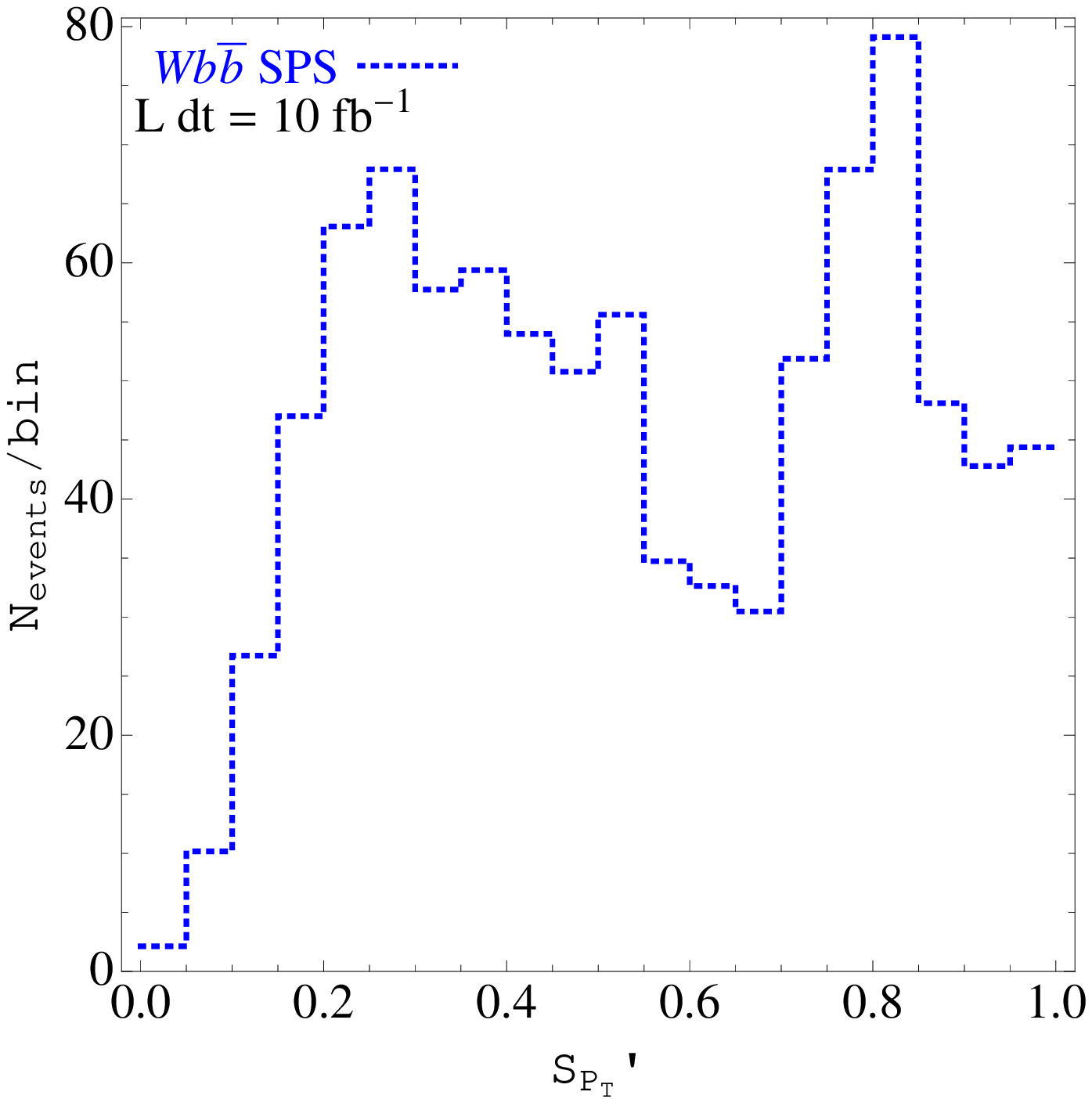}
\caption{The event rate for $Wb\bar{b}$ production from DPS (left) and SPS (right) as a function of $S_{p_T}^\prime$.}
\label{fg:sptprime-Wbb-only}
\end{figure*}

In DPS production, the bottom quarks are produced roughly back-to-back such that the vector sum of their transverse momenta tends to vanish.  Likewise, the vector sum of the lepton and neutrino momenta tends to be small (with corrections from the boosted $W^\pm$).  Thus, the $S^\prime_{p_T}$ distribution for the DPS process exhibits an enhancement  at low $S^\prime_{p_T}$, as shown in Fig.~\ref{fg:sptprime-Wbb-only}.  The peak does not occur at exactly $S_{p_T}^\prime = 0$ owing to NLO  real radiation that alters the back-to-back nature of the $b\bar{b}$ and $\ell\nu$ systems.  On the other hand, SPS production of $Wb\bar{b}$ final states does not favor back-to-back configurations, and it exhibits a peak near $S^\prime_{p_T}$ = 1.  This feature is linked to the fact that many $b\bar{b}$ pairs are produced from gluon splitting \cite{Berger:2009cm}.

\begin{figure*}[ht]
\includegraphics[width=65mm]{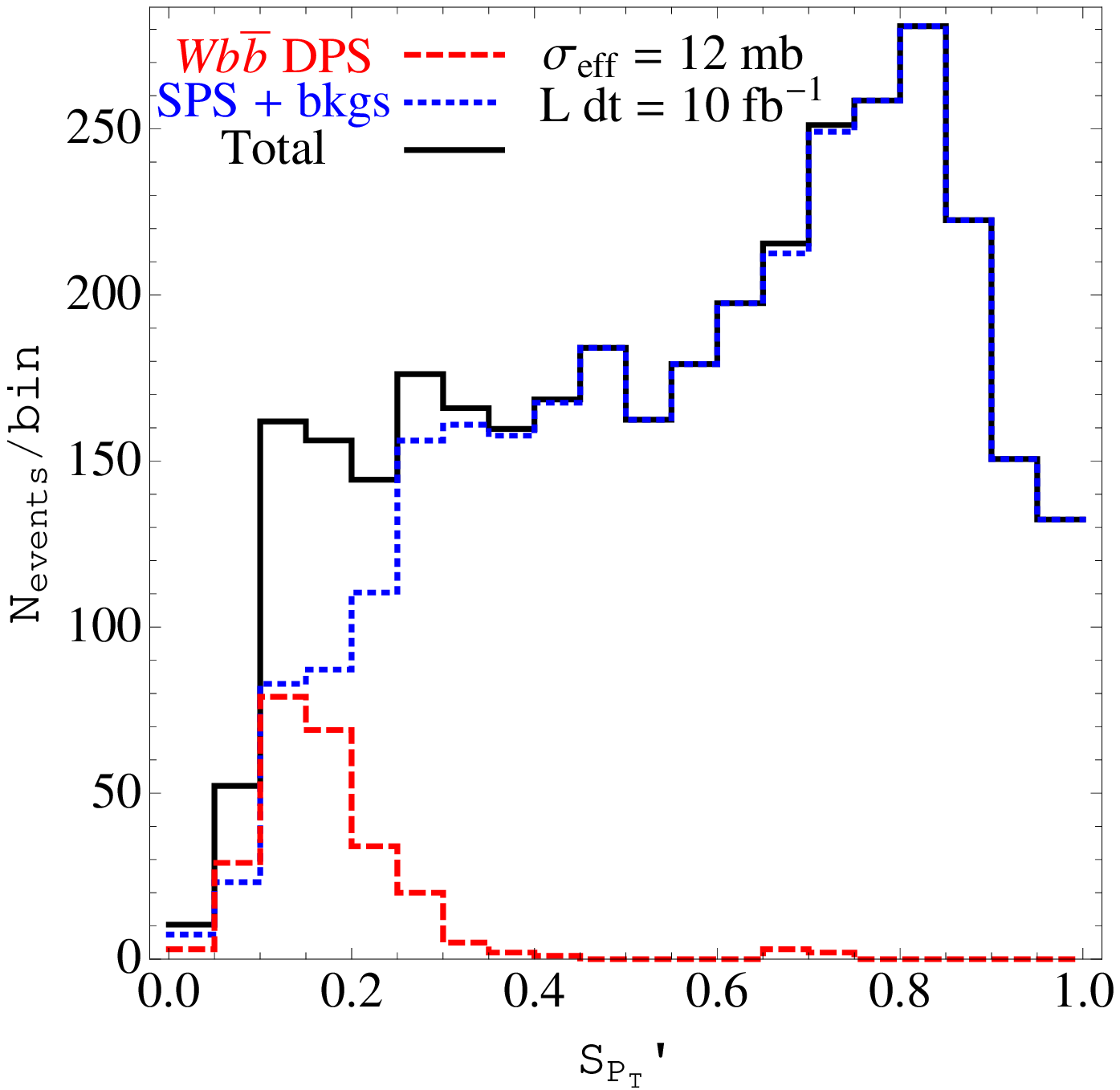}
\hspace{0.5cm}
\includegraphics[width=65mm]{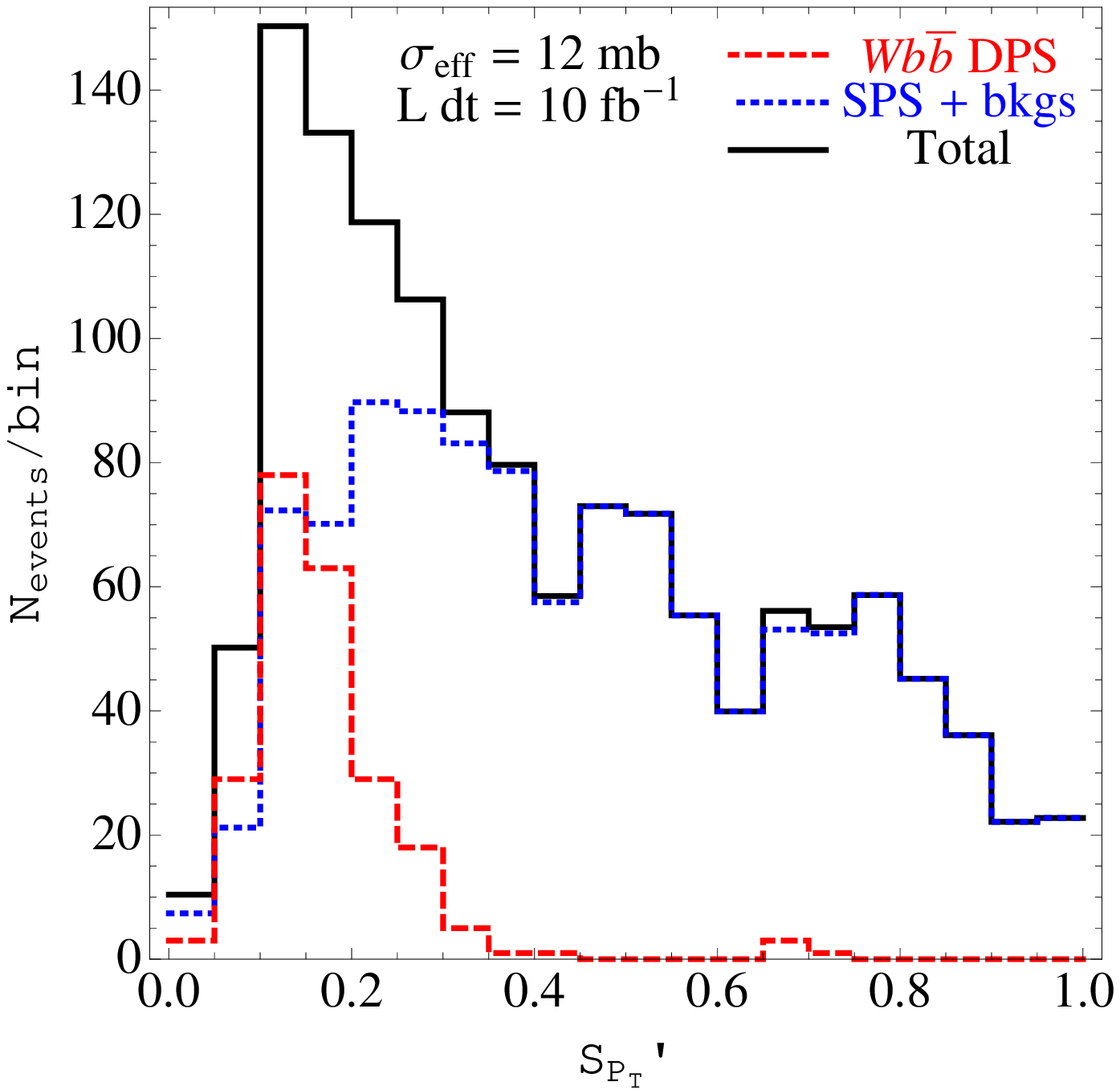}
\caption{The $S_{p_T}^\prime$ distribution for DPS and SPS production of $Wb\bar{b}$ including all relevant backgrounds.  On the left, only the minimal acceptance cuts are imposed, while, on the right, an additional maximum $\met$ cut is imposed ($\met < 45$ GeV).}
\label{fg:sptprime-w-and-wo-ETcut}
\end{figure*}

The clean separation in $S_{p_T}^\prime$ between the DPS and SPS $Wb\bar{b}$ processes exhibited in Fig.~\ref{fg:sptprime-Wbb-only} is obscured once the $t\bar{t}$ background is included (e.g., see the left side of Fig.~\ref{fg:sptprime-w-and-wo-ETcut}).  Figure~\ref{fg:sptprime-w-and-wo-ETcut} illustrates the effectiveness of the maximum $\met$ cut in reducing the $t\bar{t}$ background in the $S_{p_T}^\prime$ distribution.  After the cut, a sharp peak is evident in the region of small $S_{p_T}^\prime$ where DPS events are expected to reside.   

The plot on the right side of Fig.~\ref{fg:sptprime-w-and-wo-ETcut}, shows that extraction of a relatively clean DPS sample can be accomplished by imposing a maximum $S_{p_T}^\prime$ cut.  The last column of Table~\ref{tbl:cut-results} shows that a cut $S_{p_T}^\prime < 0.2$ reduces the SPS $Wb\bar{b}$ rate while leaving the DPS signal relatively unaffected.  In the end, the major background arises from DPS $Wjj$, as is expected since this process inhabits the same kinematic regions as the DPS $W b\bar{b}$ signal.  Despite this background, we find a statistical significance for the presence of DPS $W b \bar{b}$ of $S/\sqrt{B} = 173/\sqrt{197} = 12.3$.

\subsection{Further discrimination}
\label{subsec:other-bkgds}

\begin{figure*}[ht]
\includegraphics[width=65mm]{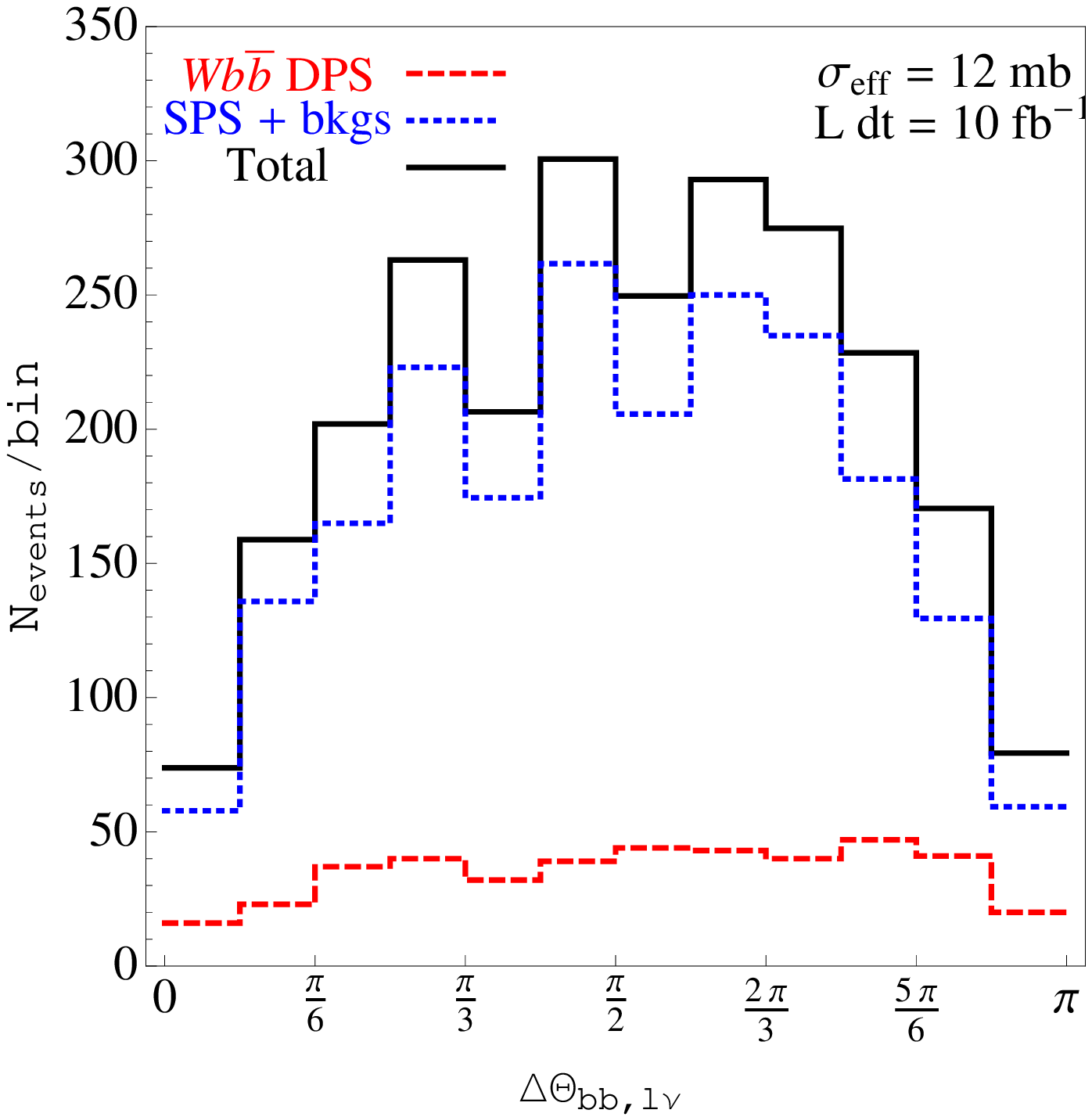}
\hspace{0.5cm}
\includegraphics[width=65mm]{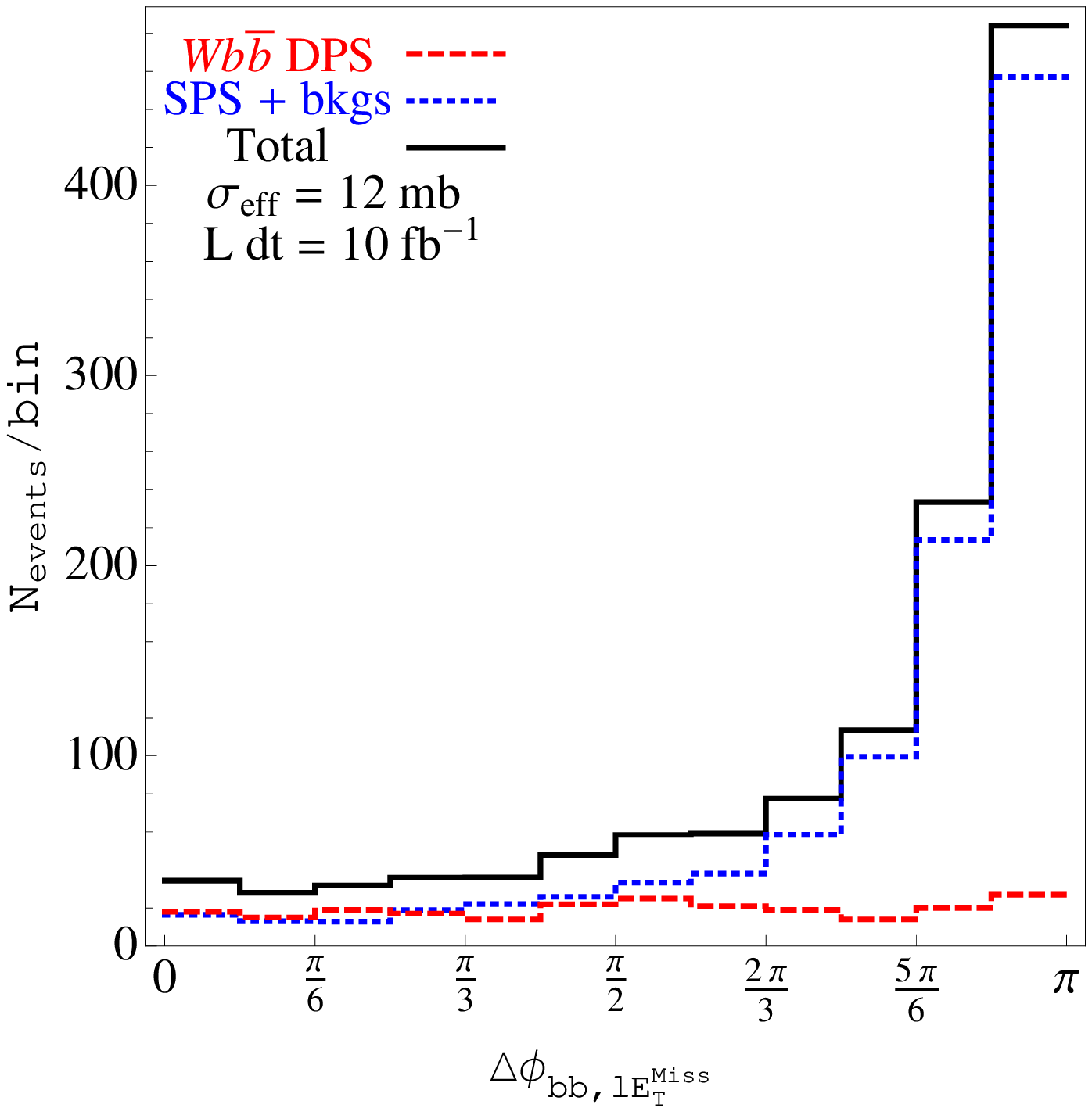}
\caption{The event rate as a function of the angle between the normals of the two planes defined by the $b\bar{b}$ and $\ell \nu$ systems (left), and the azimuthal angle between the total momentum vectors of the $b\bar{b}$ and $\ell\nu$ systems (right).}
\label{fg:delta-phi}
\end{figure*} 

Observables which take into account the angular distribution of events are also useful in the search for DPS.  In the left plot of Figure~\ref{fg:delta-phi}, we show the event rates for DPS $Wb\bar{b}$ and the backgrounds as a function of  the angle between the normals to the two planes defined by the $b\bar{b}$ and $\ell\nu$ systems.  These planes are defined in the partonic center-of-mass frame and are specified by the three-momenta of the outgoing jets or leptons.  The angle between the two planes defined by the $b\bar{b}$ and $\ell\nu$ systems is $\cos \Delta\Theta_{b \bar{b},\ell\nu} = \hat{n}_3(b_1, b_2) \cdot  \hat{n}_3(\ell,\nu)$, where $\hat{n}_3(i,j)$ is the unit three-vector normal to the plane defined by the $i-j$ system.  We require full event reconstruction using the on-shell $W$-boson mass relations to determine $\hat{n}_3(\ell,\nu)$, and use both solutions to the quadratic equation.  We see that the distribution of the DPS events is rather flat, aside from the
cut-induced suppressions at $\Theta_{b\bar b, \ell \nu} \sim 0$ and $\sim \pi$,  whereas the SPS events show a strong correlation, with a distribution that peaks near $\Delta \Theta_{b\bar{b},\ell\nu} \sim \frac{\pi}{2}$. 

In the right plot, we show the event rates as a function of the azimuthal angle between the total momentum vectors of the $b\bar{b}$ and $\ell\nu$ systems.  Since this azimuthal angle is defined in the transverse plane, it requires only $\met$, and no neutrino longitudinal momentum reconstruction.  The shape of the DPS distribution is flat while the SPS distribution shows a strong correlation, with a preference for values toward $\pi$.

We also considered the azimuthal angle between the charged lepton and the total momentum vector of the $b\bar{b}$ system, since this was less sensitive to detector effects and higher-order corrections than using the neutrino-lepton system.  However, this is far less efficient a discriminator, and for brevity, is omitted. 

In both plots of Fig.~\ref{fg:delta-phi}, it is clear that DPS and SPS exhibit different behaviors as a function of angular observables.  However, the dominance of SPS $W b \bar{b}$ and backgrounds over DPS $W b \bar{b}$ for the full range of these observables makes it impossible to extract a DPS 
$W b \bar{b}$ signal from these distributions by themselves.

\subsection{Two-dimensional distributions}
\label{subsec:2d-plots}

\begin{figure}[ht]
\includegraphics[width=80mm]{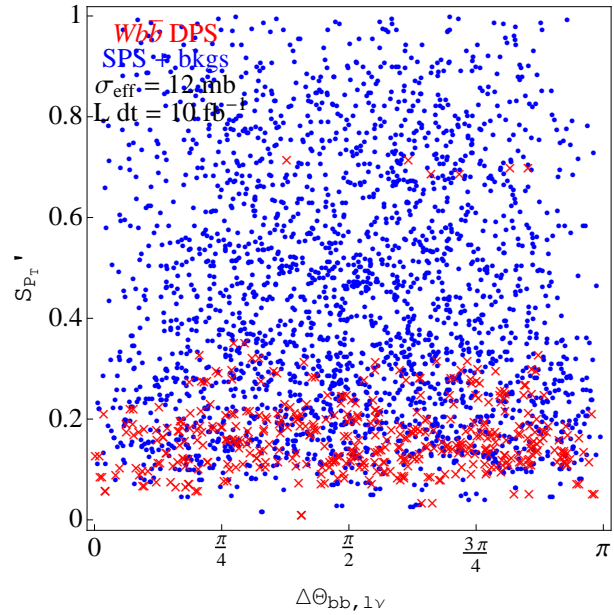}
\caption{Two-dimensional distribution of events in the variables $S_{p_T}^\prime$ and $\Delta \Theta_{b\bar{b},\ell\nu}$.  The $W b \bar{b}$ DPS events lie in the lower half of the plane, while the $W b \bar {b}$ SPS and background events occupy the upper half.  (DPS events are marked as 'x'; SPS + backgrounds with a dot.)}
\label{fg:scatter-plots} 
\end{figure} 

Despite the dominance of the $W b \bar{b}$ SPS contribution and the backgrounds over the 
DPS $W b \bar{b}$ contribution, the angular distributions can still be extremely useful when used in conjunction with other observables.  Two-dimensional distributions of one variable against another show distinct regions of DPS dominance (or SPS and background dominance).  In Fig.~\ref{fg:scatter-plots}, we construct one such scatter plot.  We show $S_{p_T}^\prime$ versus the angle between the normals of the two planes defined by the $b\bar{b}$ and $\ell\nu$ systems ($\Delta \Theta_{b\bar{b},\ell\nu}$).  We see that the DPS events reside predominantly in the lower half of the plane (small $S_{p_T}^\prime$) and are distributed evenly in the angular variable.  The separation between DPS $W b \bar{b}$ and the SPS 
component is not as pronounced in the $S_{p_T}^\prime - \Delta \Theta_{b\bar{b},\ell\nu}$ plane as we 
saw in our earlier study of $b \bar{b} j j$~\cite{Berger:2009cm}.  In the $W b \bar{b}$ case, the background events are more evenly distributed over the full plane, to some extent resulting from inclusion of both solutions for the neutrino's longitudinal momentum in the $W^\pm$ decay. 

\begin{figure*}[ht]
\includegraphics[width=65mm]{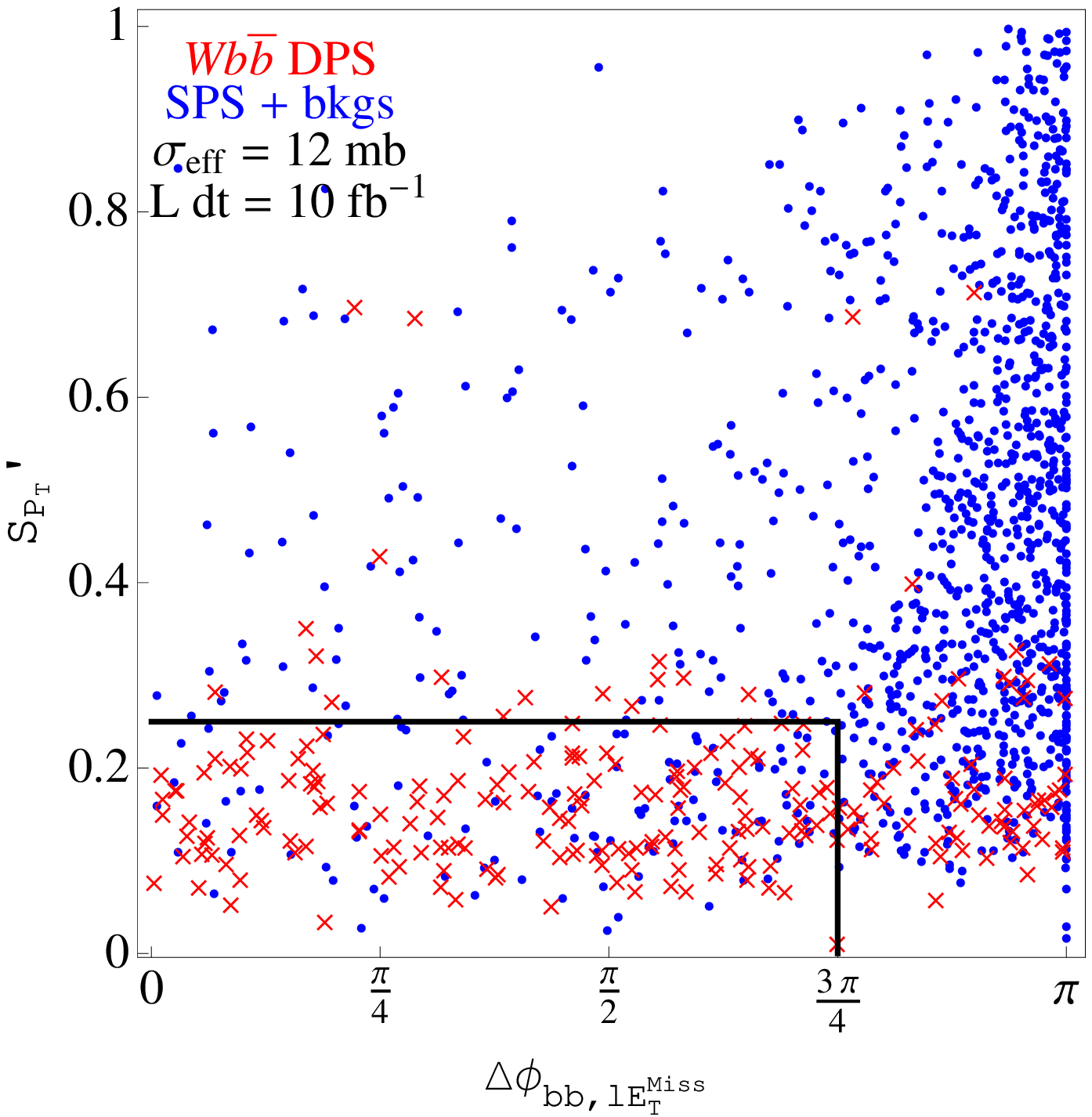}
\hspace{0.5cm}
\includegraphics[width=65mm]{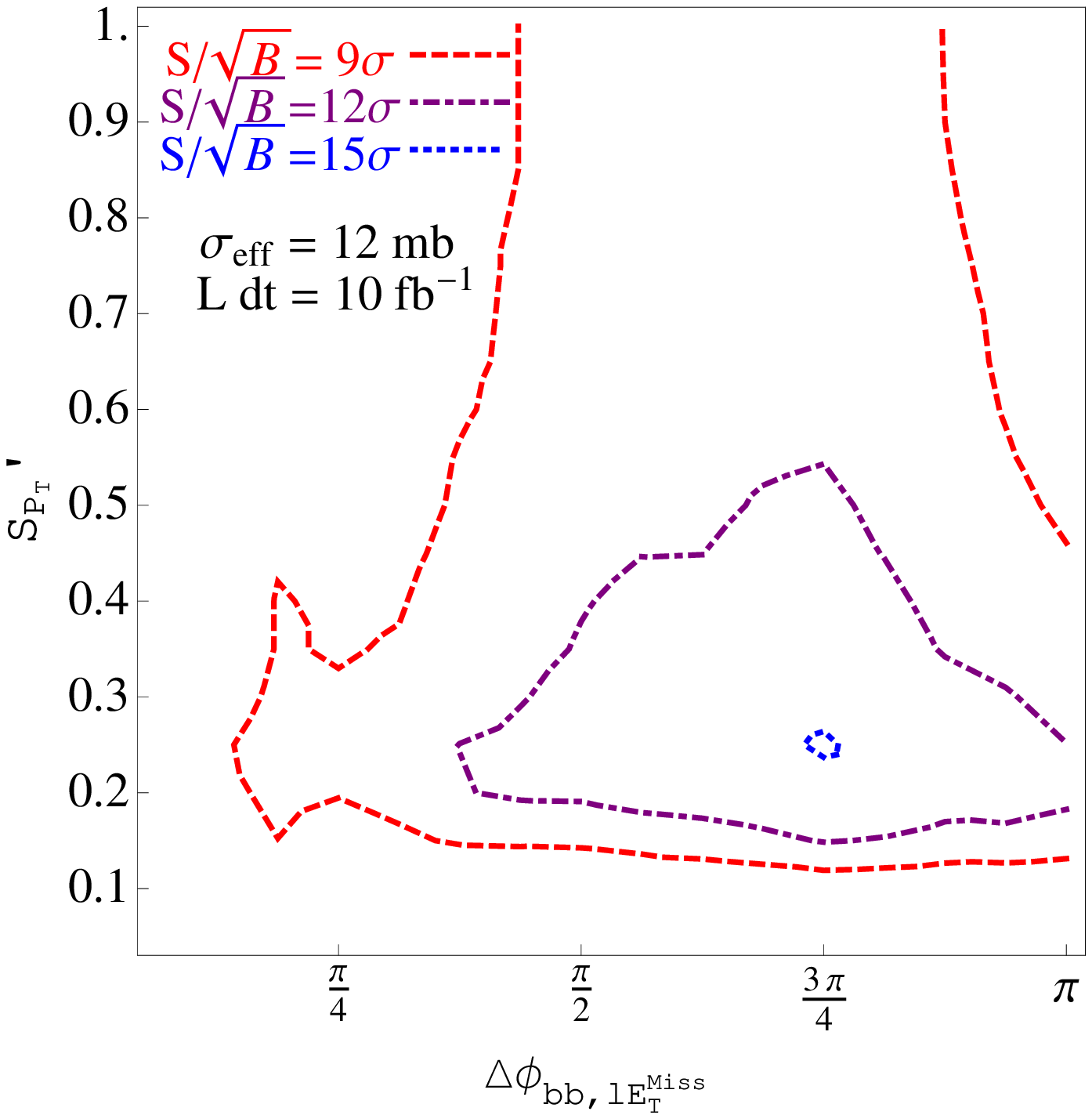}
\caption{The two-dimensional distribution of events in the variables $S_{p_T}^\prime$ and $\Delta \phi_{bb,\ell\met}$ (left).  The box denotes the boundary which gives the highest statistical significance.  On the right, we show the significance dependence as a function of the corners of the box.}
\label{fg:scatter-contour-plots}
\end{figure*}

In Fig.~\ref{fg:scatter-contour-plots}, the SPS $W b \bar{b}$ and background events in the $S_{p_T}^\prime - \Delta \phi_{b\bar{b},\ell~\met}$ plane show a strong preference for the upper right-hand corner 
of the plane.  This two-dimensional distribution indicates that cuts on the $S_{p_T}^\prime$ and $\Delta \phi_{b\bar{b},\ell~\met}$ variables should permit extraction of an enriched sample of DPS $W b \bar{b}$ events.
To quantify the degree of separation, we define a region in this plane that gives the highest statistical significance.  Its boundary is denoted by the black box in the left panel of Fig.~\ref{fg:scatter-contour-plots}.  Restricting $S_{p_T}^\prime < 0.25$ and $\Delta \phi_{bb,\ell~\met} < 3 \pi/4$, we find a statistical significance of $S/\sqrt{B} = 15.2$.  

By employing distributions in both $S_{p_T}^\prime$ and $\Delta \phi_{bb,\ell~\met}$, we achieve a better  significance than from $S_{p_T}^\prime$ alone.  In the right plot of Fig.~\ref{fg:scatter-contour-plots}, we show the dependence of the significance on the placement of the box.  As long as the maximum value of $\Delta \phi_{bb,\ell~\met}$ is in the $\pi/2-3\pi/4$ range, a statistically significant extraction of DPS $Wb\bar b$ from the other events can be obtained, given our assumed effective cross section $\sigma_{\rm eff} = 12$~mb and luminosity. 

We suggest experimental analyses of $W b \bar{b}$ at the LHC in terms of the two-dimensional distributions presented in this section with the goal to establish whether a discernible DPS signal is  
found.  The enriched DPS event sample can be used for a direct measurement of the effective cross section $\sigma_{\rm eff}$.

\section{Conclusions}
\label{sec:conclusions}

In this talk, we investigate the possibility to observe double parton scattering at the early LHC in the 
$p p \rightarrow Wb\bar{b} X \to \ell \nu b\bar{b} X$ process.  

After identifying the most relevant background processes, we pinpoint a set of observables and cuts which would allow for the best separation between the DPS $W b \bar {b}$ signal and the backgrounds (including the SPS $Wb\bar{b}$ process).  To provide the most precise predictions possible, we generate the DPS $W b \bar {b}$ signal event sample,  the SPS $W b \bar {b}$ sample, and the 
dominant  background event samples at next-to-leading order in QCD.  The main obstacles in the extraction of the DPS signal are the backgrounds from 
$t\bar{t}$ production and the SPS $Wb\bar{b}$ component.   The most efficient way to suppress the $t\bar{t}$ background is with an upper cut on the missing transverse energy of the event, since top quark decays result in larger values of $\met$.  

To separate the DPS component of $Wb\bar{b}$ from the SPS component, we find it useful to employ observables which take into account information on the full final state rather than observables which involve one or two particles.  Examples are the $S_{p_{T}}^{\prime}$ variable (defined in Eq.~(\ref{eq:Sptprime})) and the angle ($\Delta \Theta_{b\bar{b},\ell\nu}$) between the two planes defined by the $b\bar{b}$ and $\ell \nu$ systems, respectively.   By displaying the information from these two observables in two-dimensional distributions, we show in Sec.~\ref{subsec:2d-plots}
that it is possible to identify distinct regions in phase space where the DPS events reside.  Utilizing cuts on these observables that enhance the DPS $Wb\bar{b}$ sample, we find that the DPS signal can be observed with a statistical significance in the range $S/\sqrt{B} \sim 12 - 15$.

Once DPS production of 
$Wb\bar{b}$ is observed, it will be interesting to assess its potential significance as a background in searches for other physics, such as Higgs boson production in association with a $W$ boson (where the Higgs boson decays as $H \to b\bar{b}$), and precise studies of single top quark production where new physics could contribute to the $Wtb$ vertex.

\begin{acknowledgments}
This work was supported by the U.~S.\ Department of Energy under Contract No.\ DE-AC02-06CH11357.
\end{acknowledgments}

\bigskip 

\end{document}